\documentstyle[aps,preprint]{revtex}

\def\half{\frac{1}{2}}
\def\<{\left<}
\def\>{\right>}
\def\|{\left|}
\def\J{{\cal J}}

\tighten

\begin{document}

\draft

\preprint{nucl-th/9604007}
\title{Quantum Correlations in Nuclear Mean Field Theory
       Through Source Terms}

\author{S.J. Lee
 \footnote{Electronic address: ssjlee@nms.kyunghee.ac.kr
       \ and \ sjlee@ruthep.rutgers.edu}
  }
\address{Department of Physics, College of Natural Sciences \\
    Kyung Hee University, Yongin, Kyungkido  \  449-701 \  KOREA}

\maketitle

\begin{abstract}
Starting from full quantum field theory, various mean field
approaches are derived systematically.
With a full consideration of external source dependence, the stationary
phase approximation of an action gives a nuclear mean field theory
which includes quantum correlation effects (such as particle-hole
or ladder diagram) in a simpler way than the
Brueckner-Hartree-Fock approach.
Implementing further approximation, the result can be reduced
to Hartree-Fock or Hartree approximation.
The roll of the source dependence in a mean field theory is examined.
\end{abstract}



\narrowtext

One of important problems in nuclear physics is understanding
systematically the nuclear systems ranging from the structure of
stable finite nuclei to a very hot and dense system which may occur
in a high energy heavy ion collision or in a neutron star.
A nuclear system is a strongly interacting many body system
and thus the description of the system becomes complicated.
A modern approach to the study of a nuclear system is based on
the relativistic field theory in terms of relativistic nucleons
interacting each other by exchanging mesons.
The simplest approach of such a theory is the so-called relativistic
mean field approximation (RMF) or quantum hadrodynamics (QHD)
which describes a nuclear system in terms of nucleonic Dirac field
interacting with classical meson fields \cite{walecka74}.
In this approach, a nuclear system is composed of relativistic
nucleons whose self energy is determined through meson fields
which are generated by the nuclear density.

In spite of the success of this simple model in describing
various properties of nuclear system with effective interaction
\cite{walecka74,sjlthe,mill74,advnp16,sjlaxial,axial2,axial3,%
axial4,horowser}, 
it is failed to reproduce nuclear saturation properties
from the free nucleon-nucleon interaction. The calculations yield
correct binding energy with too large density or vice versa
forming the Coester band \cite{coester,kuemm}.
Even though the Dirac-Brueckner-Hartree-Fock (DBHF) approximation
including ladder diagram \cite{shakin,brock,malf,mach,shi,boers}
can achieve the nuclear saturation for a nuclear matter \cite{brock,malf},
it failed in reproducing the ground state properties for finite nuclei
using the nucleon-nucleon interaction forming a Coester band
\cite{muth,fritzmu}.
The DBHF is essentially the RHF with correlation effects (ladder
diagrams) determined through the Bethe-Salpeter (BS) equation
for a nucleon-nucleon interaction. 
Two coupled self-consistent equations, Dyson's equation and BS equation,
make the DBHF calculations very complicated for a finite nucleus
\cite{fritzmu}.
Thus we need to search for a description that is simpler than
the DBHF and can consider various quantum correlation effects.
On the other hand, various calculations \cite{mill74,fritz}
show that the exchange term, the vacuum fluctuation,
and the ladder diagrams can not be treated as perturbation
with respect to the Hartree term or each other exhibiting
non-perturbative characteristics of strong nuclear interaction.

 For a systematic study of various mean field approaches,
it is necessary to investigate the basic assumptions underlying to
each of the mean field theories starting from the quantum field theory.
In this paper, we will consider full external source dependence of
a stationary phase approximation (SPA) of the path integral for
a nuclear system with arbitrary external sources.
The added source terms bring the quantum correlations into a mean
field approach in a simple way.
Depending on the interaction considered,
this approach can give a correction to the DBHF \cite{sjlnew}
and also reveals the origin of non-perturbative
relationships among the various mean field approximations.


Up to few hundreds MeV energy, a nuclear system may be described
as a system of nucleons interacting each other via meson exchange.
 For a nuclear saturation property, we at least need to include
a scalar meson for a long range attraction and a vector meson
for a short range repulsion. 
The simplest Lagrangian describing a nuclear system can be written
in the form of
\begin{eqnarray}
 {\cal L}(x) &=& \bar\psi(x) [i \gamma_\mu \partial^\mu - M] \psi(x)
     + \half [\partial_\mu \varphi(x) \partial^\mu \varphi(x) 
                - m^2 \varphi^2(x)]
     + g \bar\psi(x) \Gamma \varphi(x) \psi(x) .  \label{lagr}
\end{eqnarray}
Here $\psi$ is the nucleon field and $\varphi$ represents meson
fields (scalar field and vector field here).
The $\Gamma$ is the unit matrix for a scalar meson
and it is the Dirac matrix $\gamma^\mu$ for a vector meson field.
More detail differences between scalar and vector meson fields
are irrelevant for the discussions in the context of this paper.

In quantum field theory, any physical observable can be represented
as a function of field operators. Introducing arbitrarily small
external sources as
\begin{eqnarray}
 {\cal L}(x,\J\!) = {\cal L}(x) -  \bar\psi(x) J(x) 
             - \bar J(x) \psi(x) - J_m\!(x) \varphi(x) ,  \label{lagj}
\end{eqnarray}                                        
the expectation value of an operator
$\hat{O} (\psi, \bar{\psi}, \varphi) $ can be obtained by
\begin{eqnarray}
 \<\!\hat{O}  (\psi, \bar\psi , \varphi) \!\>
    = \left. \frac{1}{W(\J\!)} O\!\left(\frac{i\partial}{\partial\bar{J}},
       \frac{i \partial}{\partial J} , \frac{i \partial}{\partial J_m} 
       \!\right)\! W(\J\!) \right|_{\J=0} ,   \label{expeco}
\end{eqnarray}
where the transition amplitude $W(\J)$ can be obtained through a path
integral \cite{aberlee,olando,sjlnmf};
\begin{eqnarray}
 W(\J) &=& e^{i S(\J)} 
      ~=~ \<\Psi_f\left|T[e^{-i \int H(\J) dt}]\right|\Psi_i\> 
      ~=~ \int {\cal D}(\psi,\bar\psi,\varphi) ~e^{i \int d^4 x~
            \<{\cal L}(x,\J)\>} .       \label{wj}
\end{eqnarray}                                        
Here $T$ represents time ordering
and $\J$ stands for $J$, $\bar J$, and/or $J_m$.
Notice here that $\bar J(x)$ is not the conjugate function of $J(x)$;
both are independent arbitrary functions.
If we use the single particle state representation, then the matrix
element $\<{\cal L}(x,\J)\> = \<\Psi_+\left|{\cal L}(x,\J)\right|\Psi_-\>$
becomes a functional of the single particle
wave functions $\psi(x)$ and $\varphi(x)$. 
Now the quantum field theory for a nuclear system has been reduced
to a problem of finding the corresponding transition amplitude $W(\J)$
or action $S(\J)$.

 For a Lagrangian of the form of Eqs.(\ref{lagr}) and (\ref{lagj}),
we can integrate Eq.(\ref{wj}) over the meson field $\varphi$.
However the result becomes fourth order in the nucleon field $\psi$
and thus further integral is not feasible.
On the other hand, we can integrate over the nucleon field first,
but the resulting nonlinear equation of $\varphi$ prevents further
integration.
Thus we need to use some approximate method.
However, since we can not use a perturbative method due to the
strong interaction, we are forced to use a classical or semi-classical
trajectory as the simplest nonperturbative method
\cite{sjlnew,aberlee,olando,sjlnmf,sjlbul}.
If we treat all the fields as classical ones,
then $\<{\cal L}(x,\J)\>$ in Eq.(\ref{wj})
becomes the Lagrangian ${\cal L}$ of Eq.(\ref{lagj}) with c-functions
$\psi(x)$, $\varphi(x)$, and $\J(x)$.
To treat a nuclear ground state as a Slater determinant of occupied
single nucleon levels, we should treat the nucleon field as a quantum field. 
Then the external sources $J(x)$ and $\bar J(x)$ should also be treated
as q-functions having the same quantum characteristics as the nucleon
field with infinitesimal amplitudes to keep the source terms in the
transition amplitude \cite{sjlnew}.
Similarly, $J_m(x)$ should be a q-function if we consider a quantized
meson field $\varphi(x)$.

The stationary phase approximation (SPA) or steepest descent method
of the path integral Eq.(\ref{wj})
or equivalently the variation of the Lagrangian with respect to each
field gives the corresponding Euler-Lagrange equation for a classical
trajectory as
\begin{eqnarray}
 \left[ i \gamma_\mu \partial^\mu - M + g \Gamma \varphi(x)\right]
      \psi(x) ~=~ J(x) ,     \label{nucleq}  \\
 \left[ \partial_\mu \partial^\mu + m^2 \right] \varphi(x)
    ~=~ g \bar\psi(x) \Gamma \psi(x) - J_m(x) .      \label{mesoneq}
\end{eqnarray}                                            
The equation of motion for $\bar\psi$ is given as the conjugate
of Eq.(\ref{nucleq}). 
Using the meson propagator $D(x-x')$,
we can eliminate meson fields from Eqs.(\ref{nucleq}) and
(\ref{mesoneq}). The result becomes
\begin{eqnarray}
 \left[i \gamma_\mu \partial^\mu - M + g^2 \left( \int d^4 x' 
  D(x-x') \bar\psi(x')  \Gamma \psi(x') \right) \Gamma \right] \psi(x)
        \hspace{2cm}  \nonumber  \\
     =~ J(x) +  g \Gamma \psi(x) \int d^4 x' D(x-x') J_m(x') ,
               \label{nucleonf}
\end{eqnarray}
for a ground state of a nuclear system.
This is a nonlinear equation of the nucleon field.
Since the equations of $\psi$ (Eq.(\ref{nucleonf})) and $\bar\psi$
are coupled nonlinear equations, their solutions
would be functions of infinite
order in the external source $\J$, i.e.,  $J$, $\bar J$ and $J_m$.
Usually this source dependence was neglected,
and thus the mean field theory gave the RMF of Walecka \cite{walecka74}
or the Hartree-Fock (RHF) approximation only.
Since Eq.(\ref{nucleonf}) should be satisfied order by order in $\J$,
we may expand the solution as a power series of $\J$;
\begin{eqnarray}
 \psi(x) &=& U(x)  + \psi_1(x) + \psi_2(x) 
         + \psi_3(x) + \cdots .    \label{psij}
\end{eqnarray}
Here $U(x) = \psi_0(x)$ is the source independent solution,
which is non-zero only for occupied states, and $\psi_n(x)$ is
the $n$-th order term of the external sources $J$, $\bar J$ and 
$J_m$, i.e., $\psi_n(x) = \psi(x, \J^n)$.

The source independent solution $U(x)$ can be obtained from     
Eq.(\ref{nucleonf}) by neglecting all the external
source dependent terms;
\begin{eqnarray}
 \left[ (i \gamma_\mu \partial^\mu - M) + g^2 \left( \int d^4 x' 
           D(x-x') \bar U (x') \Gamma U(x') \right) \Gamma \right] U(x) 
    ~=~ 0 .             \label{uxh}
\end{eqnarray}
Notice that the source independent field $\bar U(x)$ is the conjugate
field of $U(x)$.
If we treat meson field $\varphi$ in Eq.(\ref{nucleq}) as a classical
field, i.e., replace the right hand side of Eq.(\ref{mesoneq})
by the corresponding expectation value as it has been done by Walecka,
then $\bar{U}(x')\Gamma U(x')$ inside the integral of Eq.(\ref{uxh})
becomes a classical quantity $\<\bar{U} (x') \Gamma U(x')\>$ and
is independent of $U(x)$ appearing outside the integral.
This is the relativistic Hartree (RH) or the RMF of Walecka \cite{walecka74}.
If we treat the meson field as a quantum field, then both the
$U(x')$ and $U(x)$ are the same field and thus we have a nonlinear
equation for $U$. To find the solution $U(x)$ using
an iterative method, we can linearize this equation as
\begin{eqnarray}
 \int d^4 x' ~A(x,x') U(x') ~=~ 0 ,    \hspace{-3.2cm} \label{uxhf}  \\
 A(x,x') &=& \left[ (i \gamma_\mu \partial^\mu - M)
                  + g^2 \int d^4 x'' D(x-x'') \bar U(x'') 
                   \Gamma U(x'') \Gamma \right] \delta(x-x')
           \nonumber \\
          & & -~ g^2 \Gamma U(x) D(x-x') \bar U(x') \Gamma .
                \label{axxp}
\end{eqnarray}
Here the minus sign of the last term in Eq.(\ref{axxp})
originates from the anti-commuting property of the nucleon field.
Notice here that $A(x,x')$ is hermitian and the self-consistent
field $U(x)$ in a nuclear system is different from the Dirac field
in a free space.
The $U(x)$ and $U(x'')$ appearing in Eq.(\ref{uxhf}) through 
$A(x,x')$ of Eq.(\ref{axxp}) are now treated as independent
quantities from $U(x')$ of Eq.(\ref{uxhf})
in the iterative calculation for the solution $U(x)$.
This is the relativistic Hartree-Fock (RHF) approximation.
The second term of $A(x,x')$ is the direct Hartree contribution to
the nuclear self energy and the third term is the exchange Fock
contribution to the self energy.
In Ref.\cite{sjlthe}, it was shown that we can obtain RHF using the
quantum operator algebra without using the concept of linearlization
explicitly.

On the other hand,
from the $n$-th order terms of $\J$ in Eq.(\ref{nucleonf}),
we obtain coupled equations for $\psi_n$ and $\bar \psi_n$.
Their solutions \cite{sjlnew} are
\begin{eqnarray}
 \psi_n(x) &=& \int d^4 x' B^{-1}(x,x') J_n(x')   \nonumber  \\
     & & -~ g^2 \int d^4 x' B^{-1}(x,x') \Gamma U(x') 
         \int d^4 x_2 \int d^4 x_3 D(x'-x_3) \bar J_n(x_2) 
         A^{-1}(x_2 ,x_3) \Gamma U(x_3) ,   \label{psinx}    \\
 \bar\psi_n(x) &=& \int d^4 x' \bar J_n(x') B^{-1}(x',x)  \nonumber \\
     & & -~ g^2 \int d^4 x_2 \int d^4 x_3 \bar U(x_2) \Gamma  
         A^{-1}(x_2 ,x_3) J_n(x_3) \int d^4 x' D(x_2 -x') \bar U(x') 
         \Gamma B^{-1}(x',x) ,      \label{psibnx}
\end{eqnarray}
where
\begin{eqnarray}
 J_n(x) &=& - g^2 \sum_{n_1=0}^{n-1} \sum_{n_2=0}^{n-1}
           \sum_{n_3=0}^{n-1} \Gamma \psi_{n_1}(x) \int d^4 x' D(x-x') 
           \bar\psi_{n_2}(x') \Gamma \psi_{n_3}(x')   \nonumber \\
        & & +~ J(x) \delta_{n,1}        
            + g \Gamma \psi_{n-1} (x) \int d^4 x' D(x-x') J_m(x') ,
                   \label{jnx}      \\
 \bar J_n(x) &=& - g^2 \int d^4 x' D(x'-x) \sum_{n_1=0}^{n-1}
           \sum_{n_2=0}^{n-1} \sum_{n_3=0}^{n-1} \bar\psi_{n_1}(x') 
           \Gamma \psi_{n_2}(x') \bar\psi_{n_3}(x) \Gamma  \nonumber \\
        & & +~ \bar J(x) \delta_{n,1}        
            + g \int d^4 x' J_m(x') D(x'-x) \bar\psi_{n-1}(x) \Gamma ,
                    \label{jbnx}
\end{eqnarray}
with
\begin{eqnarray}
 n_1 + n_2 + n_3 &=& n \  ; \hspace{0.7cm}
    n_1 < n  ,  \  \  n_2 < n , \  \  n_3 < n .                               
\end{eqnarray}   
Notice here that $J_n$ and $\bar J_n$ are given in terms of $\psi_m(x)$
and $\bar\psi_m(x)$ with $m = 0$, 1, ..., $n-1$ which are lower order
in $\J$. Here
\begin{eqnarray}
 B(x,x') &=& A(x,x')             \nonumber \\
  & & -~ (g^2)^2 \Gamma U(x) \left[\int d^4 x_2 \int d^4 x_3 D(x-x_3) 
      \bar U(x_2) \Gamma A^{-1} (x_2 ,x_3) \Gamma U(x_3) \right] 
      D(x'-x_2) \bar U(x') \Gamma .      \label{bxxp}
\end{eqnarray}
If $\psi_n(x)$ and $\bar\psi_n(x)$ were independent \cite{sjlnew},
then $B(x,x')$ would be replaced simply with $A(x,x')$ and only the
first terms survive in Eqs.(\ref{psinx}) and (\ref{psibnx}).

Once we solve Eq.(\ref{uxh}) for $U(x)$, then we can find $\psi_n(x)$
and $\bar \psi_n(x)$ for all order $n$.
Now the meson field $\varphi(x)$ can be found using Eq.(\ref{mesoneq}).
Using these results, we can evaluate the path integral of Eq.(\ref{wj})
in SPA and find the corresponding action integral $S(\J)$.

Since only up to second order terms are needed in calculating
propagators of each field or the energy and the density of a system,
let's look at up to the second order in the external source $\J$.
Dropping the integral sign and the irrelevant $\J$-independent terms,
the action integral becomes up to the relevant second order in $\J$ as,
\begin{eqnarray}
 S(\J) &=& - \bar U(x) J(x) - \bar J(x) U(x)
         - \bar J(x') B^{-1}(x',x) J(x)      \nonumber  \\
    & &  \hspace{-1cm}
        - g \bar U(x) \Gamma U(x) D(x-x') J_m(x')
        + \half J_m(x) D(x-x') J_m(x')
                   \nonumber  \\
    & &  \hspace{-1cm}
        -  g^2 J_m(x) D(x-x_1) \bar U(x_1) \Gamma B^{-1}(x_1,x_2)
                 \Gamma U(x_2) D(x_2-x') J_m(x')       
                 \nonumber  \\
    & &  \hspace{-1cm}
        + \half g^4 \bar U(x_2) \Gamma A^{-1}(x_2,x_3) \Gamma U(x_3) 
            D(x_3-x_4) J_m(x_4) D(x_2-x_1)    
            \bar U(x_1) \Gamma B^{-1}(x_1,x) \Gamma U(x) D(x-x')J_m(x') 
                         \nonumber  \\
    & &  \hspace{-1cm}  
        + \half g^4 J_m(x') D(x'-x) \bar U(x) \Gamma B^{-1}(x,x_1) \Gamma 
             U(x_1) D(x_1-x_3) J_m(x_4)      
             D(x_4-x_2) \bar U(x_2) \Gamma A^{-1}(x_2,x_3) \Gamma U(x_3) .
                   \label{sj2}
\end{eqnarray}
Using Eq.(\ref{expeco}) with $W(\J) = e^{i S(\J)}$ of Eq.(\ref{sj2}),
we can find the expectation
value of any operators up to two body form.

The mean nucleon field and the mean meson field become
\begin{eqnarray}
 \<\psi(x)\> &=& \left. \frac{1}{W(\J)} \left( \frac{i \partial}
                 {\partial \bar J(x)} \right) W(\J) \right|_{\J = 0} 
             ~=~ U(x) ,          \label{meanpsi}  \\
 \<\varphi(x)\> &=& \left. \frac{1}{W(\J)} \left( \frac{i \partial}
                 {\partial J_m(x)} \right) W(\J) \right|_{\J = 0}
             ~=~ \int d^4 x' D(x-x') \bar U(x') \Gamma U(x') ,   
                  \label{meanphi}
\end{eqnarray}
as expected for a mean field theory.
The mean meson field in a nuclear system is generated as a virtual
meson due to the (real) nucleon field.
The meson field propagator becomes,
dropping the obvious integral sign,
\begin{eqnarray}
 i \Delta (x-x') &=& \<T \varphi(x) \varphi(x')\> 
               ~=~ \left. \frac{1}{W(\J)} 
                  \left( \frac{i \partial}{\partial J_m(x)} \right)   
                  \left( \frac{i \partial}{\partial J_m(x')} \right) 
                  W(\J) \right|_{\J=0}     \nonumber  \\
     &=& - i D(x-x') + g^2 D(x-x_1) \bar U(x_1) \Gamma U(x_1) 
           \bar U(x_2) \Gamma U(x_2) D(x_2 - x')   \nonumber  \\
     & & +~ 2i g^2 D(x-x_1) \bar U(x_1) \Gamma B^{-1}(x_1 , x_2) 
           \Gamma U(x_2) D(x_2 -x')     \nonumber  \\
     & & + i g^4 D(x_3-x) \bar U(x_2) \Gamma A^{-1}(x_2,x_3)
            \Gamma U(x_3) D(x_2-x_1)     
            \bar U(x_1) \Gamma B^{-1}(x_1,x_4) \Gamma U(x_4) D(x_4-x')
                         \nonumber  \\
     & & + i g^4 D(x-x_4) \bar U(x_4) \Gamma B^{-1}(x_4,x_1) \Gamma 
             U(x_1) D(x_1-x_3)           
             \bar U(x_2) \Gamma A^{-1}(x_2,x_3) \Gamma U(x_3) D(x'-x_2) ,
                       \label{deltxxp}
\end{eqnarray}
which has the free meson part and the nucleon field
dependent part similarly as in other approximations.
In RMF, only the second term appears.
On the other hand, the propagator of nucleon field becomes
\begin{eqnarray}
 i G(x,x') &=& \< T \psi(x) \bar\psi(x') \> 
         ~=~ \left. \frac{1}{W(\J)} 
             \left( \frac{i \partial}{\partial \bar J(x)} \right)  
             \left( \frac{i \partial}{\partial J(x')} \right) 
             W(\J) \right|_{\J=0}      \nonumber  \\
   &=& i B^{-1}(x,x') + T [U(x) \bar U(x')]  .       \label{gxxpg2}
\end{eqnarray}
The second term is the density dependent propagator which is also
appearing in RMF or RHF.
The first term is the correction of the Feynman
propagator $i A^{-1}(x,x')$ of the RHF.
One of the differences of our method from other mean field approaches
comes through the difference $B(x,x') - A(x,x')$ 
which corresponds to the ladder diagram included in the DBHF.

The second term of Eq.(\ref{deltxxp}) and the second term of
Eq.(\ref{gxxpg2}) form the RHF while the first term of Eq.(\ref{gxxpg2})
gives a correction to the vacuum fluctuation of the RHF.
The last three terms of Eq.(\ref{deltxxp}) gives one and two baryon
loops with the Feynman propagator $i B^{-1}(x,x')$ for one of the
internal baryon line. This propagator contains higher order
correlations (such as ladder diagram) 
which were missed in the Feynman propagator $i A^{-1}(x,x')$ of the RHF.
The particle-hole correlations can also be considered in this model
through the Feynman propagators $i A^{-1}(x,x')$ and $i B^{-1}(x,x')$
in Eqs.(\ref{sj2}) and (\ref{deltxxp}).

If we neglect the coupling term of $\psi_n(x)$ and $\bar\psi_n(x)$
\cite{sjlnew}, then there would be no higher order correlation effects 
in the Feynman propagator than RHF, i.e., we would have $B(x,x') = A(x,x')$.
If we neglect the nuclear external source, i.e., set $J(x) = 0$
in Eq.(\ref{nucleq}), then the third term of Eq.(\ref{sj2}) does not 
exist. This means that the vacuum fluctuation comes in through the $J$
dependence in a mean field theory.
On the other hand, if we neglect the mesonic external source, i.e.,
set $J_m(x) = 0$ in Eq.(\ref{mesoneq}),
then only the first four terms of Eq.(\ref{sj2}) survive.
This approximation gives the RHF with nuclear vacuum fluctuation.
If we neglect both $J(x)$ and $J_m(x)$, then we have the RHF without
vacuum fluctuation.
If we neglect any $\J$ dependence of $\psi$ in Eq.(\ref{mesoneq}),
i.e., replace $\bar\psi(x)\Gamma\psi(x)$ in Eq.(\ref{mesoneq})
by $\bar U(x)\Gamma U(x)$, then we get the RHF with the first five
terms in Eq.(\ref{sj2}). However if we use the expectation value
$\<\bar U(x) \Gamma U(x)\>$ for $\bar\psi(x)\Gamma\psi(x)$ in
Eq.(\ref{mesoneq}), then we have the RH with vacuum effect.
Various quantum correlation effects come into a mean field approach
depending on how the source dependence of fields are treated.


In conclusion, we have shown that the SPA of the full quantum field
theory with external sources gives quantum correlation effects
which were not included in other
mean field approaches. We also have seen that various assumptions
on the external source dependence reduce our SPA to RMF, RH, and RHF.
Since these further assumptions on the source dependence
are not perturbative, the various mean field approaches of
RMF, RH, RHF, DBHF, and SPA can not have perturbative relationships.
The correlation of ladder diagram is included in the SPA through
the self energy of the propagator $i B^{-1}(x,x')$ without any further
self-consistency condition in contrast to the DBHF
where the ladder diagram is included through the self-consistent
Bethe-Salpeter equation. 
Only the HF wave function $U(x)$ is required the self-consistent
calculation in the SPA in contrast to the DBHF which uses
the Dyson's equation for the nucleon propagator
and the Bethe-Salpeter equation for the effective interaction.
We should investigate further for the differences between the SPA
and the DBHF by analyzing more details and
by applying the model to a nuclear system \cite{sjlnew}.
To study n-n interaction with this model, we need to consider the
action up to 4-th order in $\J$.
We also need to extend the SPA further to the case with nonlinear
meson field to consider a nonlinear sigma model or a quark-gluon system.
This extension would enable the SPA to consider many body interaction
within the mean field theory.

This work was supported in part by the Korea Science and
Engineering Foundation under Grant No. 931-0200-035-2
and in part by the Ministry of Education, Korea through Basic Science
Research Institute Grant No. 95-2422.
I would like to thank the hospitality offered me at the Department
of Physics and Astronomy, Rutgers University.

\end{document}